\documentclass[sigconf]{acmart}

\usepackage{CJKutf8}
\pdfoutput=1

\AtBeginDocument{%
  }

\usepackage{todonotes}
\usepackage{comment}
\usepackage{graphicx}
\usepackage{lipsum}
\usepackage{float}
\setcopyright{none}
\acmConference[ACM CS \& Law Work-In-Progress Submission]{...}{2025}

\begin{document}
\pagestyle{plain}
\renewcommand\footnotetextcopyrightpermission[1]{}  

\title{Missing the Mark: Adoption of Watermarking for Generative AI Systems in Practice and Implications under the new EU AI Act}
\author{Bram Rijsbosch, Gijs van Dijck, Konrad Kollnig}
\affiliation{%
  \institution{Maastricht University - Law and Tech Lab}
  \country{The Netherlands}
}
\email{firstname.lastname@maastrichtuniversity.nl}

\renewcommand{\shortauthors}{Rijsbosch et al.}

\begin{abstract}
AI-generated images have become so good in recent years that individuals often cannot distinguish them any more from “real” images. This development, combined with the rapid spread of AI-generated content online, creates a series of societal risks. Watermarking, a technique that involves embedding information within images and other content to indicate their AI-generated nature, has emerged as a primary mechanism to address the risks posed by AI-generated content. Indeed, watermarking and AI labelling measures are now becoming a legal requirement in many jurisdictions, including under the 2024 European Union AI Act. Despite the widespread use of AI image generation systems, the practical implications and the current status of implementation of these measures remain largely unexamined.

The present paper therefore provides both an empirical and a legal analysis of these measures. In our legal analysis, we identify four categories of generative AI deployment scenarios and outline how the legal obligations could apply in each category. In our empirical analysis, we find that only a minority number of AI image generators currently implement adequate watermarking (38\%) and deep fake labelling (18\%) practices. In response, we suggest a range of avenues of how the implementation of these legally mandated techniques can be improved, and  publicly share our tooling for the detection of watermarks in images.\footnote{This paper was accepted to the 2025 ACM CS \& Law Work-in-Progress track}.
\end{abstract}


\maketitle
\section{Introduction}
The rapid advancements in generative AI technologies have made it increasingly difficult to distinguish AI-generated content from human-generated content. This poses significant risks to the integrity of our information ecosystem, democratic processes, and societal trust, especially when AI is used to create realistic images, such as deep fakes \cite{nl_ai_risks,bird2023typology,un_ai_risks}. In response to these concerns, technical solutions such as watermarking and labelling of AI-generated content are emerging as primary mechanisms to improve transparency and accountability. Indeed, several jurisdictions, most notably the European Union, have begun introducing legal obligations that require the implementation of such measures \cite{bird2023typology,eu_ai_act}. Despite growing regulatory attention, it is however unclear to what extent such fundamental technical solutions are actually used in practice and thereby can offer important protections for citizens. Such insights are especially important to uncover because laws like the 2024 EU AI Act require adequate (water)marking and labelling of AI-generated content. This paper seeks to address this gap through both a legal and an empirical analysis.

While technical solutions for marking, labelling, and detecting AI-generated images are well-developed \cite{bird2023typology,sternography_paper}, their implementation is far from straightforward. Providers of generative AI systems face conflicting incentives. From a societal perspective, the implementation of such techniques is essential to reduce risks from AI-generated content. At the same time, providers often want to offer their customers the ability to create content without any visible or invisible signs of being artificially generated. This poses additional challenges for larger companies in the AI space that both deploy generative AI systems and develop the large models behind them. Here, there also is a risk of model collapse, meaning the quality of AI models has been shown to degrade when trained on AI-generated content \cite{model_collapse}. Plus, social media companies may face liability for the distribution of content that poses societal risks under laws like the 2022 EU Digital Services Act (DSA), further providing an incentive for large providers to mark AI-generated content and limit its spread.

To align financial and societal incentives better, companies face increasing legal requirements in the AI space. The 2024 EU AI Act mandates two key measures to mitigate the risks posed by AI-generated content: (1) the embedding of machine-readable markings in all AI-generated outputs to facilitate automated detection of synthetic content and (2) visibly disclosing the artificial origin of AI-generated “deep fakes” \cite{eu_ai_act}. In case of non-compliance, organisations face high potential fines: up to 15 million Euros or 3\% of a company’s global annual turnover. These rules will start to apply from 1 August 2026, and their implementation could prove to be a foundational step towards ensuring trust and mitigating the risks of generative AI content. However, significant ambiguities exist around the practical application of these requirements. This includes questions on how to allocate responsibilities along the complex generative AI supply chain, how to assess the adequacy of existing technical solutions, and how to interpret definitions used by the AI Act (e.g., the definition of a ‘deep fake’) \cite{dewitte2024chatbots,engler2022reconciling,deepfakepaper}.

\textbf{Contributions.} In light to these uncertainties, this paper makes two contributions:
\begin{itemize}
    \item \textbf{Legal analysis and clarification of the EU AI Act:} We analyse the applicability of the AI Act’s transparency requirements in different common deployment scenarios of generative AI text-to-image systems. These insights can contribute to a successful implementation of the EU AI Act in practice and inform the design of AI content transparency measures beyond the EU.
    \item \textbf{Empirical study and technical measurement of status quo:} Taking into account our legal analysis, we provide empirical evidence on the current status quo of the implementation of machine-readable marking and visible disclosure solutions for AI-generated images. We share our methodology and tooling to detect marking and labelling, so that other researchers can further study the space (\url{github.com/BramRijsbosch/WatermarkDetectionTool}).
\end{itemize}

Since the AI Act’s transparency requirements are only applicable from August 2026 onwards, the purpose of our empirical study is not to evaluate compliance. Rather, we aim to present and discuss the current status quo of the implementation of machine-readable (water)marking and visible disclosure solutions for AI-generated image content, and the challenges associated with implementing and enforcing legal requirements (such as in the AI Act) related to such solutions.

\textbf{Structure.} The remainder of this paper is structured as follows. Section~\ref{sec:Background} reviews relevant literature on existing technical solutions for marking and labelling AI-generated image content and introduces the general framework of the EU AI Act. Section~\ref{sec:Legal-analysis} presents a legal analysis of the AI Act’s transparency provisions, focusing on how these requirements could apply in different common deployment scenarios of generative AI systems. Section~\ref{sec:Methodology} details the methodology used in the empirical analysis. Section~\ref{sec:Results} presents the results of this analysis. Finally, we discuss the implications of these findings for the (implementation of the) AI Act— and beyond.
\section{Background}
\label{sec:Background}
This section introduces the primary concepts that underpin the legal analysis of Section \ref{sec:Legal-analysis} and the empirical study of Section \ref{sec:Methodology}.

\subsection{Technical background: Techniques for marking and labelling AI-generated image content}
Methods for marking AI-generated content can be classified into two broad categories: visible and invisible methods \cite{nist_ai_report}. Both approaches can be applied either during the creation of AI-generated content or by using post-generation techniques.

\subsubsection{Visible marking methods.\\}

Visible methods include visible watermarks (e.g., logos or icons within an image), content labels embedded in an image, or separate disclosure fields that can include disclaimers and warning statements about the content \cite{nist_ai_report}. The primary goal of visible markings is to inform citizens directly about the content, which is often the end-goal of the invisible methods too \cite{nist_ai_report}.

The design and implementation of effective visible disclosures remains a complex challenge \cite{disclosureStudyUvA}, as it is difficult to guarantee that average users fully comprehend visible marks that indicate the AI-generated nature of content. Additionally, visible markings embedded in an image may face practical limitations, as they can be easily removed and can degrade the user experience of the content.

\subsubsection{Invisible marking methods.\\}

Invisible marking techniques embed machine-readable information into the image content with the aim to be interpretable by technical systems rather than humans. A range of techniques exists that aim to do so, which include machine-readable watermarking solutions, metadata embeddings, and digital fingerprinting approaches \cite{nist_ai_report}.

\textbf{Watermarking techniques.}
Machine-readable watermarking techniques add subtle perturbations to images and other content that specialised algorithms can detect to identify this content as AI-generated. These techniques can be public or private, depending on whether the detection algorithm or key is publicly shared \cite{nist_ai_report}. Common methods for images include 1) altering specific pixels in the spatial domain of an image (using for example the Least Significant Bit (LSB) watermarking technique \cite{LSB}), and 2) embedding information in a transform domain of an image (such as done in image compression). The latter is a more robust but also computationally more expensive technique \cite{DCT}. Both those methods are applied post-generation \cite{nist_ai_report,diffusion}. Watermarking can also be applied during the generation phase, such as in the case of image diffusion models, which have recently rose to prominence. Here, a watermark can be applied during generation by embedding a predefined noise pattern into the initial noise of the diffusionist model \cite{diffusion}. This technique is, however, computationally expensive and requires access to a detection algorithm to uncover the watermark \cite{nist_ai_report,content_credentials_verify}.

Several developers are starting to use invisible machine-based watermarking solutions in their image generation models. Meta claims to use an invisible watermark (next to visible marking and metadata embeddings) that is applied using a deep learning model, but they do not share the corresponding model that can be used to detect this watermark \cite{meta_ai_updates_2023}. Meta has also worked with other researchers on watermarking solutions that can be embedded during the image generation process, which could prove valuable for open-source models \cite{fernandez2023stable,meta_labeling_ai_images}. Google adds an invisible watermark into the pixels of AI-generated images from their Imagen models \cite{adobe_firefly_api}, while also offering a related open-source detection tool \cite{vertex_ai_watermark}. The open-source model developers Stability AI and Black Forest Labs (together with their partners) also integrate invisible watermarking techniques in (some of) their models \cite{stability_ai_safety,stabilityai_invisible_watermark_detection}. They use an open-source watermarking library that includes both a watermarking encoding and decoder algorithm \cite{stability_ai_invisible_watermark,shieldmnt_invisible_watermark,flux_github}. Additionally, several organisations are offering API-based solutions or open-source tools that allow others to embed post-generation invisible watermarks to their images \cite{imatag_forensic_watermarking,huggingface_truepic_watermark}.

\textbf{Metadata-based marking.}
The data of an image consists of several blocks of information, with the most important being the block containing the pixels of an image. Other blocks can contain additional (textual) information, which is referred to as ‘image metadata’. There exist many standards to embed metadata into images, of which the EXIF, XMP and IPTC standards are the most used ones. Information about the AI-origin of an image can be embedded in the image using the textual formats of these standards. In the case of the IPTC standard, the ‘creator’ or ‘contributor’ field can, for example, be used to include the AI generation tool that was used \cite{IPTC2024}. The effectiveness of metadata for marking AI-generated images can be questioned, as metadata can be easily manipulated or removed from a file, and is often automatically stripped during sharing on social media platforms (to preserve bandwidth and privacy).

To prevent manipulations of metadata, it is possible to add a cryptographic signature to the metadata, so that when an image is shared, others can verify its authenticity using this signature \cite{nist_ai_report}. A prominent example is the technical specification that was developed by an industry-led consortium called the Coalition on Content Provenance and Authenticity (C2PA) to track the origin and authenticity of digital content \cite{c2pa_specification,nist_ai_report}. The C2PA specification allows for the integration of metadata embeddings, (visible and invisible) watermarking techniques, cryptographic signing options and digital fingerprinting \cite{nist_ai_report}. C2PA thereby offers an online open-source detection tool that uses trust list and cryptographic keys to store and detect the signed metadata and watermarks of images from trusted users of the C2PA standard \cite{content_credentials_verify}. In the context of metadata, this approach is again mainly relevant to verify authenticity, as metadata can easily be stripped from image files. However, when combined with watermarking and digital fingerprinting techniques, such a solution forms a more robust approach. Yet, the success of this approach will ultimately depend on the participation of the many different parties in the digital ecosystem.

\textbf{Digital fingerprinting.}
Digital fingerprints are unique identifiers that can be derived from the content of an image, such as patterns in pixel values \cite{nist_ai_report}. Fingerprinting therefore does not involve manipulating the image content, but instead relies on algorithms to generate representations (also called hashing) of the image that can be matched against external databases \cite{nist_ai_report}. An example of a digital fingerprinting solution used in practice is Youtube’s Content ID system. This system scans for copyrighted materials used on Youtube, by linking representations to an external database of verified content \cite{youtube_content_id}. Similarly, the C2PA framework also allows for the use of digital fingerprinting techniques. Using the C2PA standard, cryptographically signed hashes can be included in the metadata of an image and then stored and detected via the C2PA implementation tool \cite{content_credentials_verify}. This is currently used by Adobe \cite{pai_adobe_case_study}. Plus, stripped image metadata of a generated image can also be recovered using a hash algorithm or an embedded watermark \cite{cai_open_source_faqs,digimarc_c2pa}. Another approach to digital fingerprinting is to embed fingerprints in the training data content, which is also shown to allow detection after generation \cite{fingerprintindata}.

\subsection{Legal background: Approach and scope of the EU AI Act}

The EU AI Act was adopted in 2024 to foster safe and responsible AI development and deployment in the EU. Although other jurisdictions, such as South Korea \cite{sk_ai_legislation} and China \cite{cac2025measures}, have also introduced AI laws that contain relevant transparency rules for generative AI content, the EU AI Act is the most comprehensive and wide-ranging regulatory framework for AI to date and is therefore the reference for this paper.

The AI Act adopts a risk-based approach, whereby legal obligations depend on the risks and impact of the various AI risk categories \cite{eu_ai_act}. There are five risk categories, namely:

\begin{enumerate}
\item A limited set of AI systems with unacceptable risks, the use of which is forbidden in the EU.
\item A list of high-risk AI systems, which must adhere to strict legal requirements before they can be sold, used or made available on the EU market.
\item AI-applications with specific transparency risks (which can also include high-risk systems) that must adhere to certain specific transparency-related requirements. These include the watermarking and labelling requirements for generative AI systems.
\item In addition to AI systems, the scope of the AI Act includes rules for general-purpose AI (GPAI) models. These are models that can be integrated into a large variety of downstream applications. The models behind generative AI (image) applications are stated as a typical example of a GPAI model. Providers of such models must provide documentation and information to downstream providers. A sub-category of GPAI models — those defined as GPAI models with systemic risks — are subject to additional obligations, including the assessment and mitigation of possible systemic risks.
\item All other AI systems and/or models fall into the minimal - or no - risk category and do not face any obligations from the AI Act (other EU laws might still apply).
\end{enumerate}

The AI Act’s scope applies to both public and private entities, including non-EU actors offering or using their AI systems (or output of their systems) in the EU. Exceptions exist for research, military, and in some cases free/open-source AI systems, though the transparency requirements for generative AI systems still apply to the latter.

The AI Act distinguishes between providers (those developing and/or marketing AI systems or GPAI models under their own name/trademark) and deployers (those deploying AI systems to end-users under their own authority). Most obligations fall onto providers, with deployers facing some limited requirements in the case of deploying high-risk systems or certain AI systems with transparency risks. End-users (individuals or organisations that interact with — or are affected by — an AI system) do not face any legal obligations. Most of the AI Act’s provisions, including the transparency requirements, take effect on August 1, 2026, with fines up to €15 million, or 3\% of annual global turnover in case of non-compliance with the transparency requirements.
The AI Act distinguishes between providers (those developing and/or marketing AI systems or GPAI models under their own name/trademark) and deployers (natural or legal persons using an AI system under their own authority, except where this use is for a personal non-professional activity. Most obligations fall onto providers, with deployers facing some limited requirements in the case of deploying high-risk systems or certain AI systems with transparency risks. Most of the AI Act’s provisions, including the transparency requirements, take effect on August 1, 2026, with fines up to €15 million, or 3\% of annual global turnover in case of non-compliance with the transparency requirements.

With the AI Act just having been adopted, there is a strong need to clarify how to translate its legal requirements into technical artefacts, as aimed at by this work.
\section{Legal analysis: Watermarking obligations under the EU AI Act}
\label{sec:Legal-analysis}
Section 3.1 outlines the relevant provisions for marking and labelling of AI-generated content as set out in Article 50 of the AI Act. While these requirements may appear straightforward, their practical application can prove difficult due to the complex supply chains behind generative AI systems. To address these complexities, Section 3.2 introduces four common deployment scenarios of generative AI systems and analyses how the requirements could apply in each scenario. These scenarios subsequently serve as our basis for the empirical study of Section \ref{sec:Methodology}. 

\subsection{The AI Act’s rules for watermarking and labelling AI-generated content}
Article 50 of the AI Act includes the legal requirements for AI-applications with specific transparency risks \cite{eu_ai_act}. This includes two specific rules that apply to the providers and deployers generative AI (image) systems:

Firstly, Article 50(2) states that \textit{providers} of generative AI systems shall ensure that ``the outputs of these systems are marked in a machine-readable format and detectable as artificially generated or manipulated''. Hereby also stating that ``providers shall ensure their technical solutions are effective, interoperable, robust and reliable as far as this is technically feasible'', considering the type of content, costs of implementation and technical state of the art (Article 50(2)). The recitals list different techniques and methods that can be used or combined to fulfil this requirement, which include ``watermarks, metadata identifications, cryptographic methods for proving provenance and authenticity of content, logging methods, fingerprints or other techniques, as may be appropriate'' (Recital 133).

Further to this technical marking requirement, Article 50(4) mandates that \textit{deployers} of generative AI systems that ``generate or manipulate image, audio or video content constituting a deep fake have to disclose that the content has been artificially generated or manipulated''. Article 3(60) thereby defines a deep fake as an ``AI-generated or manipulated image, audio or video content that resembles existing persons, objects, places, entities or events and would falsely appear to a person to be authentic or truthful''. However, what constitutes as a deep fake under this definition is still open to interpretation and far from a trivial task \cite{deepfakepaper}.

Moreover, the different methods that can be used for implementing this disclosure are not specified. Article 50(5) however does state that the information referred to in both these transparency rules ``shall be provided to the natural persons concerned in a clear and distinguishable manner at the latest at the time of the first interaction or exposure''. For this study, we interpret this as a requirement to include a visible watermark, label or disclosure message directly in the output of a generative AI system, instead of adding a separate disclosure message that is not included in the image (e.g., a text label in the interface of the system). This ensures that the marking remains visible when the image is shared without any manipulations.

The transparency rules of Article 50 apply only to AI systems (e.g., apps or web tools), meaning that the developers and providers of the AI models behind generative AI applications are not required to implement any of these requirements. As also written in the AI Act, “AI models, including General-Purpose AI models, require the addition of further components, such as for example a user interface, to become AI systems” (Recital 97). This is commonly achieved through a web interface, a mobile application, or a desktop client. 

The requirements for the separate category of GPAI-models do not include any measures directly aimed at AI-generated content. The adopted Code of Practice for GPAI models does, however, currently include a commitment for signatories of the Code to use methods such as watermarks and digital fingerprinting for post-market monitoring of systemic risks and the tracing of incidents related to the use of their model \cite{gpai_code_practice_2023}.

Important to note is that in contrast to some other AI Act requirements, providers of AI systems released under free and open-source licenses are not exempt from the transparency requirements of Article 50 (Article 2(12)). Moreover, the AI Act transparency provisions are likely to apply to a large number of non-EU providers too, as most publicly available generative AI apps and websites can and are being used within the EU (Article 2(1)).

\subsection{How the rules apply in practice}
\begin{figure*}[t]
    \centering
    \includegraphics[width=0.75\linewidth]{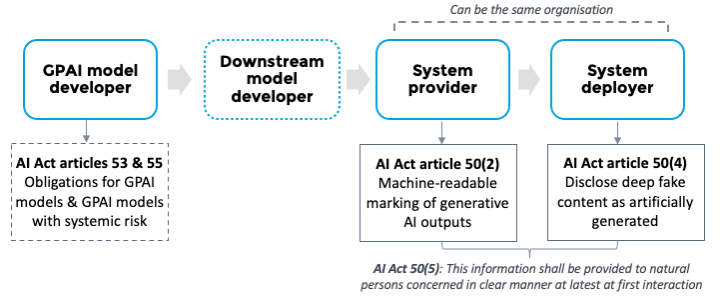}
    \caption{The applicability of the AI Act’s (water)marking and disclosure rules along a simplified version of the generative AI supply chain.}
    \label{fig:requirementsSupplyChain1}
\end{figure*}

Generative AI applications, including image generators, are the product of a complex and diverse supply chain, iinvolving a wide range of different actors \cite{dewitte2024chatbots,lee2023ai_generation}. A simplified version of the supply chain for generative AI image supply chain can be described as consisting of four primary players, each performing different functions: 1) base model developers, 2) downstream model developers, 3) system (interface) providers, and 4) system deployers. Figure \ref{fig:requirementsSupplyChain1} illustrates how the AI Act’s rules would apply along this simplified supply chain. 

In practice, however, the distinction between model developers (or model providers), system providers, and and deployers is less clear-cut. Consequently,  and the implementation of the AI Act’s transparency requirements measures is are therefore not strictly divided between these roles and remains subject to interpretation. The European Commission is expected to publish guidelines to further clarify the definitions and scope of the obligations in article 50 \cite{ec2025b_guidelines}.

One example of this ambiguity is the use of the term deployer in the deep fake disclosure requirement \cite{barbera2024ai}. The AI Act defines deployers as those \textit{using} an AI system under their own authority (except when this use is in the case of personal non-professional activities). Under this definition, deployers could be interpreted as the ‘end-users’ of generative AI systems, who query a system to generate content that is used in a professional activity (e.g., organisations generating AI content for a marketing campaign), which then requires a deep fake disclosure. This interpretation would align with terminology used in related EU product regulations, such as the Machinery Directive, which uses the term \textit{professional user} to refer to a natural person operating a machinery product in a professional context \cite{eu2023_machinery}, or the Medical Device Regulations, where \textit{user} refers to health care professionals or laypeople using the device \cite{eu2017_medical_device, eu2017_invitro}.

However, given that the AI Act does not use or define a similar a term, and considering the nature of AI systems, and how the term deployer is used in the high-risk provisions, deployer could also be interpreted as to refer to the entity that \textit{deploys} a generative AI system under its authority to end-users (e.g., via an app or website). In that case, such an entity also uses an existing system under its authority, although the nature of this usage is obviously different from that of end-users. Given the potential risks associated with generative AI systems, and the likely enforcement challenges for the disclosure requirement, it would arguably be more practical to implement disclosure measures at the system level (such as automatically applied labels within an image, or by giving end-users the option to add a label), rather than expecting professional end-users to apply such labels manually after image generation. Accordingly, and for the purposes of this study, we adopt the latter interpretation when referring to the deployer role. 

To further illustrate the challenges of applying the AI Act’s transparency requirements in practice, we identify four common deployment scenarios of generative AI systems and analyse how the relevant rules may apply in each case. These four scenarios thereby serve as the foundation for the measurement study in Sections \ref{sec:Methodology} and \ref{sec:Results}.\\

\noindent\textbf{\textit{Category 1. End-to-end integrated systems.\\}}
Most generative AI image applications depend on large-scale pre-trained base models (GPAI models), which can subsequently be fine-tuned for specific tasks or domains. The process of training such a base model is costly, requiring time, expertise, data storage, and computing power. Following the creation of a base model, a model developer often decides to use the model in their own application or platforms (directly accessible to end-users), thereby becoming the sole participant in the supply chain outlined in Figure 1. A notable example is OpenAI’s development and subsequent use of the DALL-E-3 image model in their ChatGPT application.

These organisations thus become both the provider and deployer of the generative AI system under our interpretation of the AI Act, which requires compliance with both transparency requirements. The dual role of model and system developer would allow such organisations to implement robust watermarking techniques in the model development phase (such as embedding initial noise patterns in diffusion-based models).

These end-to-end developers frequently (also) offer access to their models to other developers of generative AI applications. Access can be offered through paid channels, such as by providing networked access to their base models through an Application Programming Interface (API) \cite{dewitte2024chatbots}. This allows developers to retain control over the model even when integrated in clients’ services. Examples include OpenAI or Adobe offering API access to their image models, while also deploying these in their own application \cite{adobe_firefly_api,openai_dalle3_api}. Organisations can also adopt an (partly) open-source approach: making their models (or model weights) freely available for download and re-use by others, for example through the popular AI repository platform Hugging Face (see also category 3). An example is StabilityAI, which deploys a generative AI system via their own DreamStudio tool \cite{dreamstudio_beta}, while also offering open-source access to most of their models \cite{open_washing}.

Some organisations in this end-to-end category also operate large (social) media or image platforms (e.g., Google, X), which requires these providers to comply with the systemic risk mitigation rules under the EU’s Digital Services Act, including measures linked to generative AI \cite{guidelines_dsa}. Furthermore, certain GPAI models integrated in these end-to-end systems may fall under the AI Act’s category of GPAI models with systemic risk, which would require the implementation of additional systemic risk mitigation measures.\\

\noindent\textbf{\textit{Category 2. Systems using API model access.\\}}
Many generative AI image systems are built by leveraging (easy-to-use) APIs from large-scale GPAI model providers and integrating these APIs into user-friendly interfaces, such as mobile apps or web tools, hereby creating the generative AI system as defined under the AI Act. This approach allows organisations to use high-quality models without needing the expertise or infrastructure to build and run them. Developers of such systems often also offer additional functionalities to end-users, like the option to choose between different models or the addition of system prompts to help users improve their creations. 

Organisations leveraging models via APIs would then likely be considered both the provider and deployer of the generative AI system and, thus, will need to comply with both transparency requirements. Compliance can be achieved by relying on features (such as invisible watermarks) that are already built into the model or offered by model providers as an API-parameter option. Otherwise, organisations would have to implement their own measures after the images are created via the API. For example, through the addition of post-processing watermarks, or the embedding of metadata. Applying visible disclosure measures solely to \textit{deep fake} images could thereby provide a challenge (especially for smaller-scale organisations), as this would require a separate (NLP-based) solution capable of specifically detecting deep fake prompts. A simpler approach could involve applying visible disclosures to \textit{all} generated images, although this may negatively impact the user experience and business model of a provider.

In some cases, providers of model APIs can also offer pre-built code or templates for creating a user interface alongside the API. These model providers could then potentially be considered as system providers under the AI Act, thus bearing responsibility for the machine-readable marking requirement. However, as guidance is still to be created on the distinction between providers and deployers, we do not further analyse this case in this study.\\

\noindent\textbf{\textit{Category 3. (Open-source) Systems deployed on Hugging Face.\\}}
By providing a platform and tools for storing, testing, sharing and deploying AI models, the Hugging Face has an important role in the generative AI ecosystem. Many AI image model developers share their open-source models on the Hugging Face platform and thereby allow others to download, use and potentially fine-tune these models for specific applications. 

\begin{figure}
    \centering
    \includegraphics[width=1\linewidth]{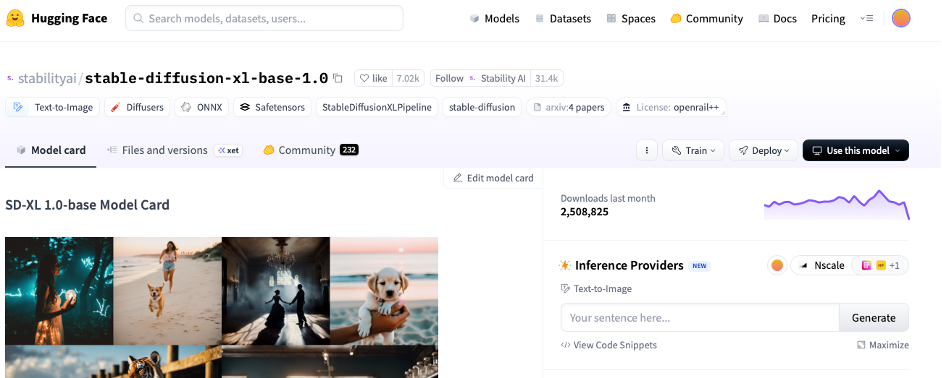}
    \caption{Example of Hugging Face’s Inference widget (bottom right), as found on the popular stable-diffusion XL-1.0 model page. This widget allows end-users to directly generate images from a model; it is yet unclear who in this case bears responsibility for implementing the AI Act’s transparency requirements.}
    \label{fig:widget}
\end{figure}

Hugging Face thereby also offers so-called interactive API inference widgets, which are pre-configured interfaces added to a model page on Hugging Face that directly allow end-users to use a model to generate images, as shown in Figure \ref{fig:widget} \cite{huggingface_serverlessinference, huggingface_inference}. Hugging Face works with inference providers who provide the infrastructure for this tool, and model owners can adjust the settings of these widgets.
Next to the API-widgets, Hugging Face also offers ``Spaces'', which is an easy-to-use deployment tool on Hugging Face that allows providers to host and offer machine learning apps to end-users. Many (small-scale) developers use Hugging Face Spaces to deploy generative AI image systems based on open-source models hosted on Hugging Face.

We include this category because in terms of the AI Act, it is not trivial to decide who would bear the responsibility for complying with the AI Act’s requirements in these cases. It could be argued that Hugging Face or its inference providers, who offer the platform and infrastructure, but have little control over the models or API’s, would be the provider and/or deployer in some of these cases. We also include this category to study whether content of popular open-source image models, like the models from Stability AI, also includes pre-baked watermarking solutions when generated via such Hugging Face tools.\\

\noindent\textbf{\textit{Category 4. Systems using other (open-source) models under their own trademark.\\}}
The final category includes organisations that use (finetuned) AI models of other organisations in their own system and deploy them under their own name and trademark (via apps or websites), without disclosing the underlying models used. Organisations in this category would then again likely classify as both the provider and deployer of a generative AI system, requiring compliance with both transparency obligations. We include this category as there are many websites and mobile apps, often listed in non-EU countries, that offer generative AI systems that fall under this category. 

Providers of such systems could thereby rely on downloaded versions of open-source models, which allows for finetuning and possibly also sharing or selling the models further (via APIs or open-source repository platforms),  which does require infrastructure to host and run these models. Or they can use API model access without disclosing or crediting the provider of the API. While API license terms typically require crediting the model developer, exceptions exist, or organisations may disregard the license terms.

The next section explores whether and how machine-readable marking and deep fake labelling solutions are currently being implemented in practice, using the four deployment scenarios outlined above as the basis for this analysis. In practice, however, distinguishing between these categories can prove challenging, due to the many ways in which generative AI models and systems are developed and shared. 
\section{Methodology: watermarking in practice}
\label{sec:Methodology}
This section describes the methodology used to analyse the current status quo in the implementation of machine-readable marking and visible disclosure measures for AI-generated content transparency. In line with good scientific practice, we publicly share the code for easily running our technical inspection images for indications of an AI-generated nature. 

\subsection{Selection of AI systems for analysis}
For our study, we aimed to select 50 of the most used generative AI systems. As explained in the following, we selected these systems from a range of different business models and distribution channels to represent a diverse range of popular tools across the four different deployment categories of Section 3.2.

Category 1 systems (end-to-end integrated) were selected by using the search query ``AI image generation'' in Google Search and identifying organisations that offer free image generation tools using their own foundational models from the top results. We only included systems that offered free text-to-image generation in Europe (if needed: by creating an account or starting a free trial).

The systems from categories 2 (using API model access) and 4 (using others’ AI models under own trademark) were similarly selected by using the search query ``AI image generation'' in the Apple App Store and Google Search. The top 15 systems from each modality (30 in total) were included. Duplicates from Category 1, such as ChatGPT, were excluded. The deployment characteristics - such as model or API mentions, or other attributions - were reviewed in documentation and interface environments to classify the systems as Category 2 or 4.

Category 3 systems (open-source systems hosted on Hugging Face) were selected by filtering the Hugging Face 'model' section on the five most downloaded open-source text-to-image generation models that offered the Hugging Face API widget tool (Figure \ref{fig:widget}) for image generation on their respective model pages \cite{huggingface_models_text_to_image}. Additionally, five of the most 'trending' Hugging Face Spaces (environment to host AI apps) for image generation were included in the analysis \cite{huggingface_spaces_text_to_image}. 

This resulted in a total of 50 systems. For each system, we also recorded general information, such as the modality, provider, location of the provider, and the AI model(s) used (if disclosed). This information is included in the Table of Appendix \ref{sec:Appendix}.  

\subsection{Generation of images with selected systems}
For each system, at least two images were generated using the standard settings. We used a neutral prompt: ``A PhD student'' and a potentially risky and clear deep fake prompt: ``A beautiful deep fake photograph of Donald Trump in the McDonald's'', to analyse if AI labels are added specifically to deep fake images. In the few cases that the prompt of Donald Trump was rejected, we used an alternative – but clear – and potentially harmful deep fake prompt, namely: ``A beautiful deep fake photograph of the river Maas in Maastricht overflowing by a flood''.

For systems that allow users to choose between different models in the systems interface, images were generated for several different available models (randomly selected). All generated images were stored using the ``save'' or ``download'' function added in the system’s interface, or otherwise using the standard save function of the phone or laptop. The images are archived on GitHub for reference.

\subsection{Detection of watermarks and disclosures}
To analyse the adoption of marking and disclosure measures in practice (in the context of the deployment categories defined in our legal analysis), we used the two evaluation metrics outlined in Table \ref{tab:metrics}.
\begin{table*}
    \centering
    \includegraphics[width=0.95\linewidth]{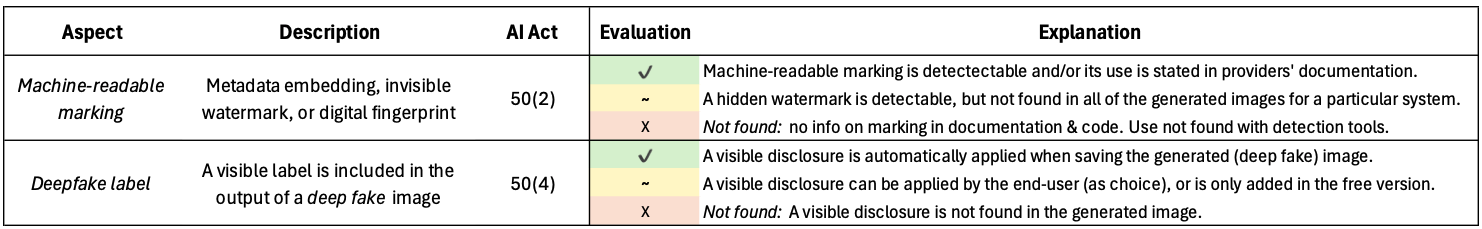}
    \caption{Metrics used for evaluating the use of marking and labelling measures in AI systems}
    \label{tab:metrics}
\end{table*}

Evaluating the implementation of the machine-readable marking requirement was done through a combination of inspecting, for each AI system, its online documentation, code (if available) and generated images (hereby relying on the review of marking techniques from Section 2.3):
\begin{itemize}
    \item \textbf{Online Documentation of AI Systems:} To meet the AI Act’s machine-readable marking requirement, information on the use of such measures must also be clearly presented to users (see Section 3.1). Therefore, each system’s online documentation was reviewed to analyse if information was provided on the use of machine-readable marking solutions. Sources included system descriptions, FAQs, terms of use, privacy policies, and ReadMe files.
    \item \textbf{Code behind AI Systems:} For systems using open-source models, the model information pages and source code (if available) were inspected for mentions of watermarking (libraries), detection tools and/or metadata adaptions.
    \item \textbf{Images Generated by AI Systems:} The following components were used to analyse the generated images for indications of having been AI-generated. These components were combined in a single tool (as shared on GitHub) that generates a Word report with the results for each of the images:
    \begin{itemize}
        \item \textit{Metadata:} The EXIF metadata extraction tool of Phil Harvey was used to analyse the metadata of the images \cite{exiftool}. This tool retrieves the metadata from all primary metadata standards (e.g., EXIF, IPTC, and XMP). The resulting metadata manually inspected for mentions of the AI-generated nature of the image, such as the use of the ‘contributor’ field in the IPTC metadata standard.
        \item \textit{Watermarking:} Identifying the use of invisible watermarks required an iterative process. If mentions of watermarking solutions were found in either the documentation or code analysis of the systems, we applied the corresponding detection tool to all the 50 images (if available). We found only two instances of invisible watermarking solutions. These are from Google \cite{vertex_ai_watermark} and the open-source watermark library used by Stability AI and Black Forest Labs \cite{stability_ai_invisible_watermark,shieldmnt_invisible_watermark,flux_github}. In both cases, a corresponding detection algorithm was also shared, which we integrated in our tool. 
        \item \textit{Digital Fingerprinting:} Verifying the use of digital fingerprinting solutions was done in a similar, iterative, manner. In this case, we could only find the C2PA technical specification (discussed in Section 2.3) being openly shared as a digital fingerprinting solution used in practice \cite{c2pa_specification}. We integrated the C2PA verification tool (also hosted on the C2PA website) in our code to check for the presence of a C2PA digital fingerprint, metadata embedding and/or invisible watermark in each of the generated images \cite{content_credentials_verify}.
    \end{itemize}
\end{itemize}

Additionally, manual inspections were carried out to verify if the generated images included any visible watermarks or other disclosure methods that could indicate that the content was AI-generated. This was done both for the neutral prompt images and the images generated with the clear deep fake prompt, as in the context of the AI Act, the requirement for visible markings only applies to deep fake images. We however did not evaluate whether the images generated with the deep fake prompt would in practice also classify as deep fakes under the AI Act’s definition of a deep fake (``falsely appearing to a person to be authentic or truthful'', Article 2(60)), as we believe more guidance is needed to perform this assessment in the context of the AI Act. 
\section{Results}
\label{sec:Results}
\begin{figure*}[h!]
    \centering
    \includegraphics[width=0.83\linewidth]{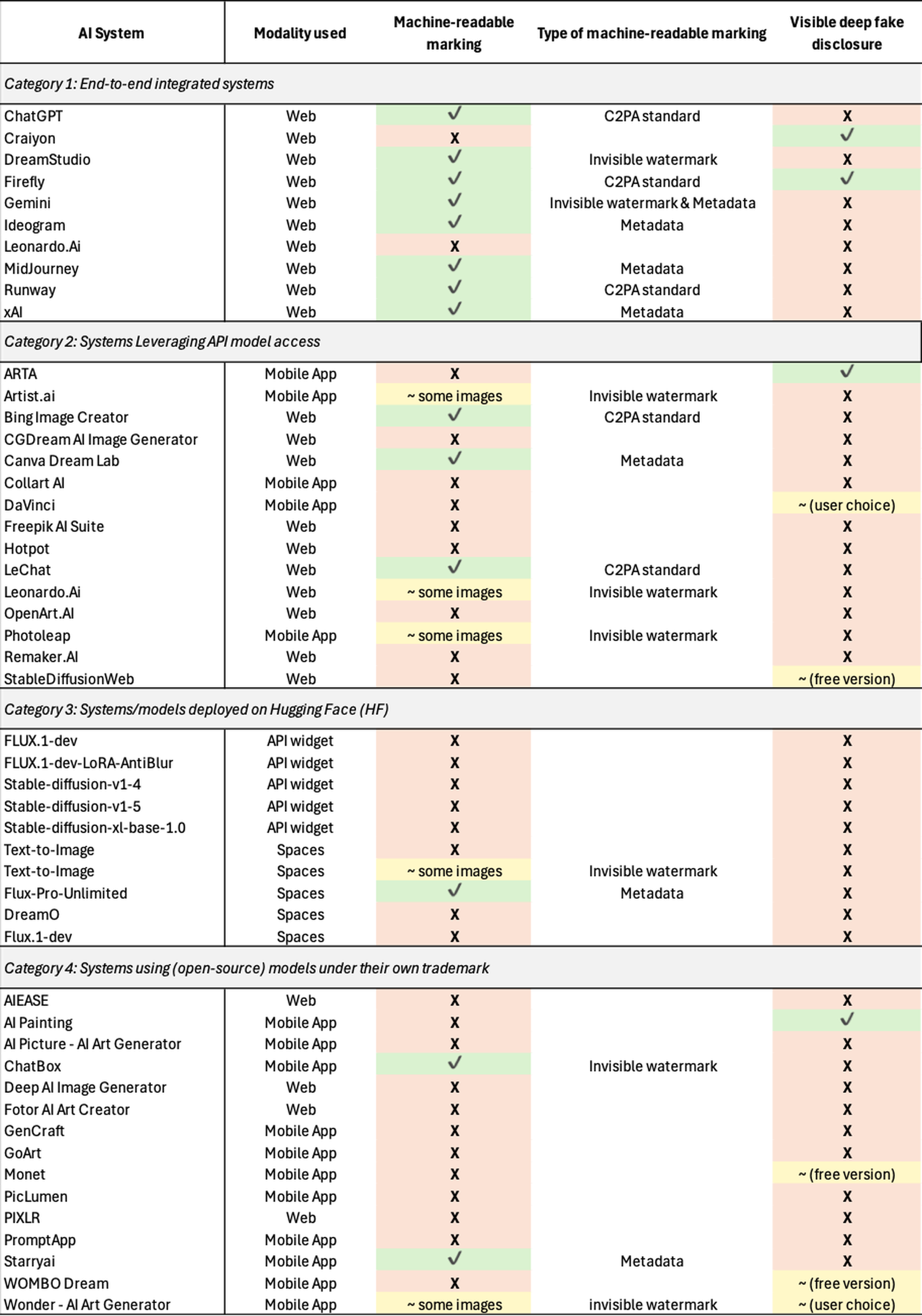}
    \caption{Results from analysing the 50 generative AI image systems, using the metrics from Table \ref{tab:metrics}}
    \label{fig:results2}
\end{figure*}

Our results are shown in Table \ref{fig:results2}, with the main findings detailed in the following paragraphs, hereby using the the four generative AI deployment scenarios of Section 3.3. 

\textbf{Limited implementation of machine-readable watermarking solutions, primarily in end-to-end systems (category 1)}. Among the 50 systems analysed, we found that 19 of these included some form of a machine-readable marking in the generated image (38\%). These solutions are mainly used by end-to-end providers from category 1 (8/10 systems) and large-scale system providers of category 2, such as Bing and Canva. Many of these providers also operate a social/digital media platform, or search engine (e.g., Google, X, Adobe, Canva, Microsoft).

\begin{figure}
    \centering
    \includegraphics[width=1\linewidth]{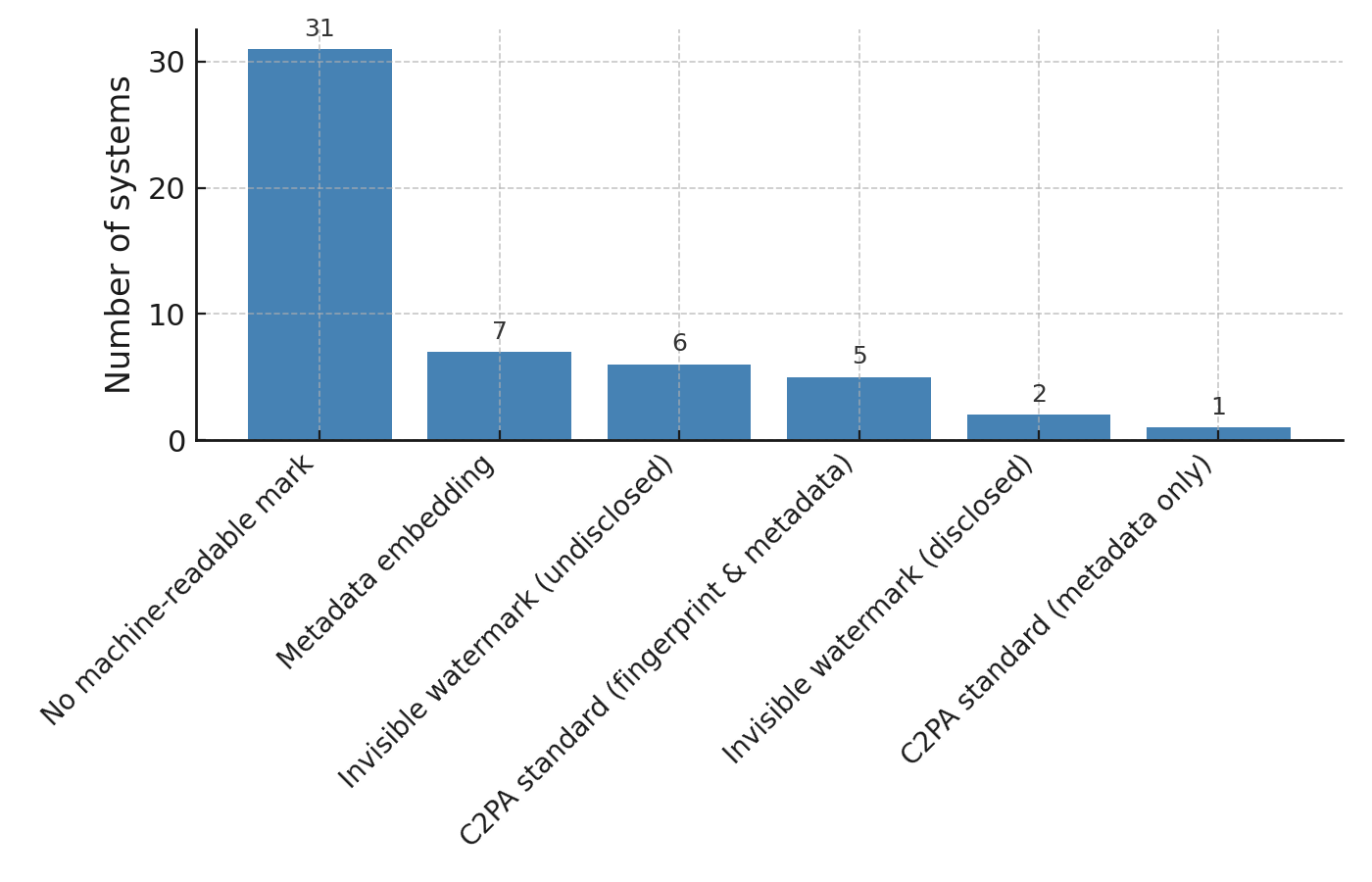}
    \caption{Overview of the different machine-readable marking solutions found in the 50 AI systems analysed.}
    \label{fig:markings}
\end{figure}

Figure \ref{fig:markings} shows an overview of the machine-readable marking solutions that were found in our analysis. The most used solution was the embedding of metadata that indicates the AI-generated origin of the data (12 systems), which is easily removable from the images. In five of these instances, the embedded metadata was part of the C2PA standard. Four out of the five providers using this standard also embedded a digital fingerprint in the image, so that the C2PA verify tool can be used to inspect the provenance of the image. Invisible watermarking techniques were found in only a small subset of the systems (8 systems). Two of these belonged to the end-to-end system providers Google and StabilityAI, as also indicated in their documentation. The other six instances of invisible watermarks were not disclosed by the providers but were found using the open-source python watermarking library that is used in the popular open-source models of Stability AI and Black Forest Labs. However, in many of the other systems that rely on the models (or APIs) from Stability AI or Black Forest Labs, we could not detect this watermark. 

\begin{figure}
    \centering
    \includegraphics[width=1\linewidth]{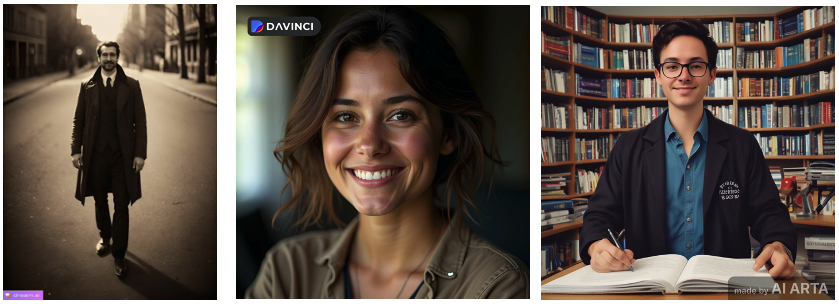}
    \caption{Examples of visible markings applied in the content generated with the neutral prompt ('A PhD student')}
    \label{fig:visiblemarks}
\end{figure}

\begin{figure}
    \centering
    \includegraphics[width=1\linewidth]{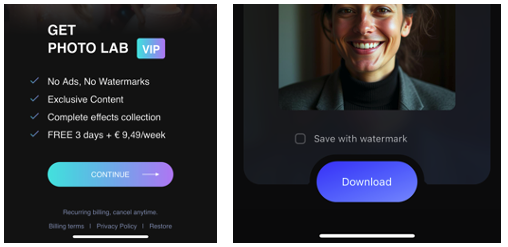}
    \caption{Examples of systems that allow end-users to remove visible markings from the generated images.}
    \label{fig:removemarks}
\end{figure}

\textbf{Visible markings for deep fake images are rarely used in practice}. Visible watermarks or other disclosure measures that are embedded within the image or AI system and that can help indicate the AI-generated nature of the content were found in only 9 out of the 50 systems (18\%). Figure \ref{fig:visiblemarks} shows three examples of such visible disclosures in images generated with the neutral prompt. In all nine cases, these markings were not exclusively applied to deep fake images, but rather to all the images generated with the system. In five out of the nine systems that used visible markings, there was an option for end-users to either add the visible marking through a checkbox, or remove the markings by signing-up to a paid plan, as illustrated in Figure \ref{fig:removemarks}. 

Interestingly, while end-to-end providers (category 1) have almost all implemented a form of machine-readable marking, visible disclosures were found to be used in only 2 of the 10 end-to-end systems. Some of these end-to-end systems, however, did restrict the use of clear deep fake prompts (e.g., Donald Trump), showing an alternative approach to tackling the risks of deep fakes.

\begin{figure*}
    \centering
    \includegraphics[width=0.95\linewidth]{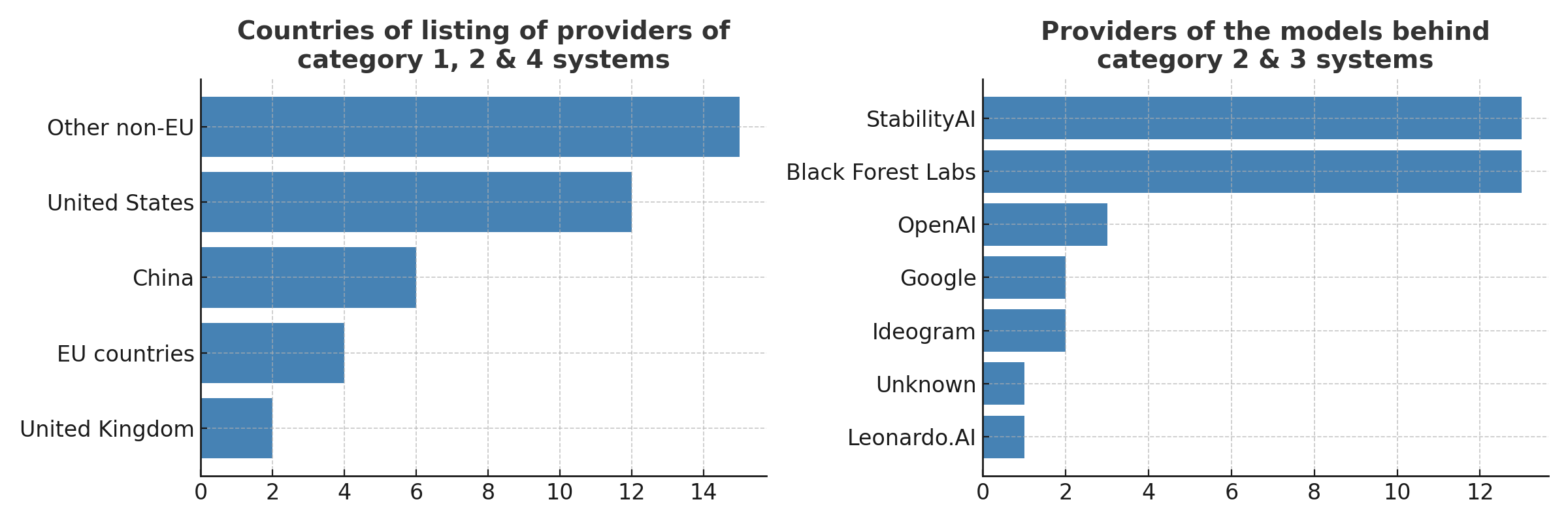}
    \caption{Overview of the main countries of listing of the system providers (left) and the different GPAI models that were indicated to be used within the systems (right).}
    \label{fig:models_used}
\end{figure*}

\textbf{Only a few different (open-source) GPAI models are used by non end-to-end providers}. We found only a limited number of GPAI models being used by the non-end-to-end systems that indicated the use of another model (i.e., category 2 and 3 systems). Almost all these models - or finetuned versions of these models - came from the popular open-source model providers Stability AI (such as their Stable Diffusion or SDXL models) and Black Forest Labs (Flux models), as illustrated in Figure \ref{fig:models_used}.

\textbf{Most providers of popular generative AI systems are based in non-EU countries}. We selected the systems of category 1, 2 and 4 based on top search results from both Google Search and the Apple App Store. We found that only in 3 out of these 40 systems the provider is listed in an EU country, as illustrated in Figure \ref{fig:models_used}.

\textbf{Watermarking solutions used by end-to-end model providers are not automatically extended to API-based systems}. Although some end-to-end operators, such as StabilityAI or OpenAI, implement machine-readable watermarks in their own models and/or systems, these features are not automatically extended to, or used by, the providers of systems that leverage those models via APIs. This is most clearly seen in the case of the models deployed on Hugging Face (category 3 systems). Here we could detect only two systems using a watermark solution, whereas the developers of the models that are being used often do implement such solutions in their own platforms or models. The same however holds for systems in category 2. Here, we do see that some providers embed metadata information or use the C2PA standard, but these embeddings are either different from the ones used by the model developers, or do not indicate the use of another base model.
\section{Limitations}
It is essential to acknowledge several limitations and constraints in our empirical analysis. First, there is a possibility that we failed to identify certain watermarking and digital fingerprinting techniques that are used in practice. We tried our best to find any disclosures of watermarking from the documentation of the providers of the systems, but there is a chance that they might have not disclosed them and used non-standard techniques that we did not check for. Second, this study has a temporal limitation. The generative AI systems and machine-readable marking solutions that we examined in this study are regularly updated, and the implementation of marking and labelling practices is likely to change over time. Third, although we used prompts that we clearly see as deep fakes, we admit that the definition of a deep fake might be subject to debate. Finally, some AI systems that we analysed did not offer a save or download button for the images within the systems’ interface environment, which required us to use other downloading methods. This may have caused certain metadata to be lost, potentially affecting our findings regarding the embedding of machine-readable watermarks.
\section{Discussion: The wild west of AI image generation}
\label{sec:Discussion}
\textbf{Markings and labels are rarely used in practice.} Our findings show that only a minority of providers currently implement machine-readable marking practices, as will soon be mandated by the EU AI Act. It is predominantly large and well-funded organisations (mostly from category 1) that implement such solutions. Given that many of these organisations also operate digital and social media platforms, the implementation of machine-readable marking measures may primarily be driven by aligning incentives, such as preventing the pollution of their (social) media platforms or AI training datasets, or protecting IP rights. 

Visible disclosures indicating the AI-generated nature of images are even more rarely used, especially among end-to-end operators from category 1. This is likely driven by the impact that such measures have on the user experience, and not because of costs or technical difficulties. At the same time, we see that several smaller organisations do use visible markings to get users to sign-up to a paid plan that allows the removal of the markings. When visible disclosures are used, they are applied to all generated images, rather than specifically deep fake images. To restrict the use of visible disclosures to specifically deep fake images, instead of all AI-generated images, providers would need a separate system - likely NLP-based - that can classify certain prompts as deep fakes, which can prove difficult and computationally expensive for smaller organisation to implement. We do, however, observe that similar methods are used in practice, as several large organisations restrict specific deep fake prompts (e.g., Donald Trump). Whether visible markings will become more widely implemented as the AI Act’s transparency requirements take effect in August 2026 remains uncertain, especially considering that most of the providers of popular generative AI systems are not based in the EU. Moreover, it is necessary to also consider the question of how effective some of the current visible markings (as seen in Figure \ref{fig:visiblemarks}) are in helping the normal layperson understand that these images are AI-generated.

\textbf{Concentrated AI ecosystem, with just a few players.} Our findings highlight the important role that providers of the most advanced (open-source) models play in shaping the ecosystem. A large number of system providers rely on models (or fine-tuned versions) from just a handful of GPAI model providers (e.g., Stability AI, Black Forest Labs and OpenAI). These providers do try to incorporate robust watermarking solutions in their models, and we find several instances of such solutions in systems relying on these models. However, these solutions are also quite easy to disable (e.g., by commenting out a line of code) and not consistently applied to API-based deployment methods. If these providers can design more robust watermarking solutions that are not easily bypassed, this could shift the paradigm, requiring model deployers to actively remove watermarks, rather than add them. Platforms such as Hugging Face can play an important role in enforcing such measures for open-source models that are hosted on their platform. Especially as Hugging Face might also qualify as a provider (and possibly deployer) of generative AI systems, when the Hugging Face widgets are used to deploy a model on the Hugging Face platform. By not exempting open-source systems from the transparency requirements, the AI Act has thus taken an important step in to address the critical role of open-source providers.

\textbf{Lack of robust machine-readable marking, even among large providers.} Given that the AI Act requires robust machine-readable marking solutions (as far as technically feasible), it is questionable if most of the currently used techniques comply with this requirement. This is particularly the case for providers that exclusively embed metadata (6 systems), which can be easily removed. Robust watermarking methods that focus on detecting AI generated images, remain rare, even among large providers. Additionally, the majority of machine-readable marking solutions are currently relying on post-generation techniques. Using robust marking methods directly in the model generation phase, particularly by category 1 GPAI model providers, could significantly ease implementation for downstream system providers using APIs and require significant extra effort for bad actors to remove. The absence of such pre-embedded solutions now forces smaller system providers to either implement cheaper and weaker methods (metadata), or to refrain from implementing any solutions at all, as observed in many cases in this study.
\section{Conclusions and how to move beyond the status quo}
In this paper, first, we studied the implications of the EU AI Act on the need to implement watermarking techniques for AI-generated images and deep fakes. We identified four categories of AI system that are relevant under the AI Act and outlined how the legal obligations could apply for each category. Next, we studied the status quo in implementing watermarking in practice, looking at 50 of the most widely used AI systems for image generation. We found that and only few providers currently implement robust watermarking solutions and that visible AI deepfake labels are yet rarely used in practice.

Our legal analysis highlights the ambiguities around the distinctions made in the AI Act between providers, deployers, models and systems in the context of the generative AI image systems, and the considerable impact that further guidance on these distinctions can have. One example is the interpretation of the term deployer in the context of the deep fake disclosure requirement. Deployer can either refer to (professional) end-users of a system, or to entities ‘deploying’ a system to such end-users. Here, it would be most practical and impactful to implement disclosure measures at the system level (such as giving users an option to add a label), rather than having end-users manually search and apply such labels themselves after generation. 

Similarly, the burden for compliance with the machine-readable marking measure may disproportionally fall on small-scale app and website providers, rather than on the large-scale and well-funded organisations that develop the models behind these applications. Yet, robust marking and labelling solutions would be best implemented at the model development stage, where they could also be enforced via API and licensing terms. Plus, enforcing compliance at the model developer level would be easier given the relatively small number of organisations that can develop advanced image models. One potential solution could be to designate the most advanced image models as GPAI models with systemic risk, which would require developers to take measures for mitigating systemic risks, such as the use of robust marking measures for deep fake images. Alternatively, it could be considered to use the implementing guidelines of the AI Act to classify providers that offer ready-to-use model APIs as providers of AI systems, as it could be argued that such an API is a way of adding an interface to a model and creating a system. This would require model API providers to implement robust machine-readable marking solutions in the model stage.

Finally, given the variety of marking techniques that are used in practice, and the growing number of deployed AI image generation systems (often deployed from non-EU countries), automated methods for compliance inspection will be essential to ensure compliance and effective enforcement when the AI Act’s requirements come into effect.

Moving forward, society is faced with an uphill battle where it will be challenging to have adequate watermarks in all AI-generated images. After all, much of the relevant AI technology is available open-source and free of charge. With models getting smaller, it will increasingly become easier to produce AI-generated images at home. However, it is reassuring that the largest providers generative AI systems do already provide protections. It will, however, be extremely difficult to protect society from actors with malicious intents in challenging our perceptions of truth and reality. In addition, while this study focused on images, the AI Act also requires the implementation of adequate marking and labelling solutions in video, audio and text generation systems, for which the implementation challenges could prove even harder. 

\bibliographystyle{ACM-Reference-Format}
\bibliography{bib/sample-base}

\appendix
\section{Appendix}
Table \ref{tab:appendix} provides additional background information on the 50 generative AI systems included in the empirical analysis.
\label{sec:Appendix}
\begin{table*}
    \centering
    \includegraphics[width=0.85\linewidth]{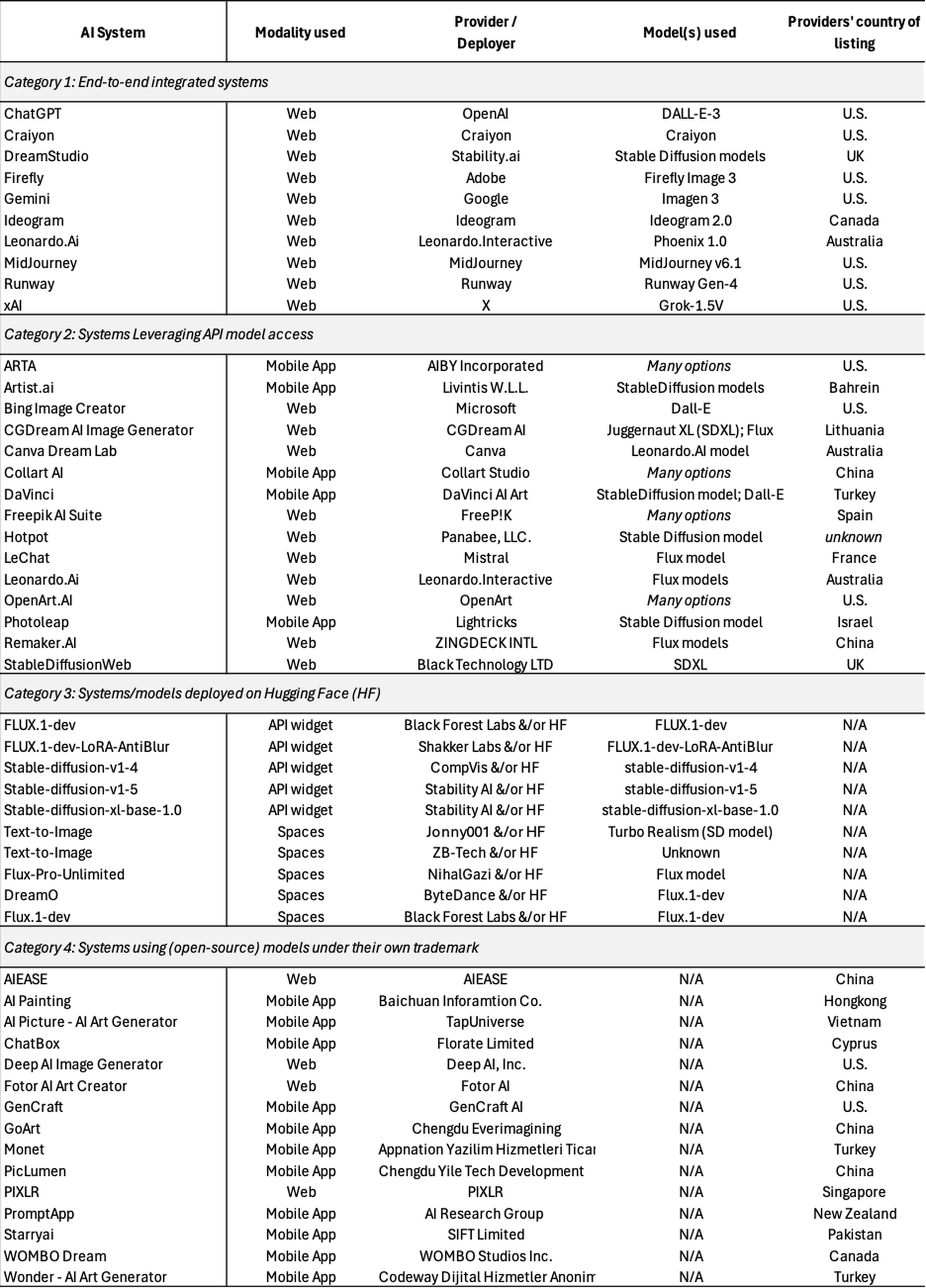}
    \caption{Overview of general information on the 50 generative AI systems that were analysed}
    \label{tab:appendix}
\end{table*}
\end{document}